# Optical and mechanical mode tuning in an optomechanical crystal with light-induced thermal effects

D. Navarro-Urrios,[1,2,a)] J. Gomis-Bresco,[1] N. E. Capuj,[3] F. Alzina,[1] A. Griol,[4] D. Puerto,[4,b)] A. Martínez,[4] and C. M. Sotomayor-Torres[1,5]

[1] *Catalan Institute of Nanoscience and Nanotechnology, Campus UAB, Edifici ICN2, 08193 Bellaterra, Spain*

[2] *NEST, Istituto Nanoscienze – CNR and Scuola Normale Superiore, Piazza San Silvestro 12, Pisa, I-56127*

[3] *Depto. Física. Universidad de la Laguna. 38206. 8Spain*

[4] *Nanophotonics Technology Center, Universitat Politècnica de València, Valencia Spain*

[5] *Catalan Institution for Research and Advanced Studies (ICREA), 08010 Barcelona, Spain.*

We report on the modification of the optical and mechanical properties of a silicon 1D optomechanical crystal cavity due to thermo-optic effects in a high phonon/photon population regime. The cavity heats up due to light absorption in a way that shifts the optical modes towards longer wavelengths and the mechanical modes to lower frequencies. By combining the experimental optical results with finite-difference time-domain simulations we establish a direct relation between the observed wavelength drift and the actual effective temperature increase of the cavity. By assuming that the Young's modulus decreases accordingly to the temperature increase, we find a good agreement between the mechanical mode drift predicted using a finite element method and the experimental one.

___________________________

[a)] Author to whom correspondence should be addressed. Electronic mail: daniel.navarrourrios@nano.cnr.it.

[b)] Present address: Laboratoire Lasers Plasmas et Procédés Photoniques (LP3), CNRS et de l Université de la Méditerranée, 163 Avenue e Luminy, Marseille FR-13288, France

## I. INTRODUCTION

The field of optomechanics has emerged as a new platform for light–matter interactions, paving the way to the implementation of micro or even nano-optomechanical systems (MOMs/NOMs) in integrated circuits [1]. Functionalities such as ultrasensitive detection of small displacements or weights and possible uses in quantum information processing are some of the appealing applications driving the fast developments in this area. Great advances in these directions have been made and recently it has been reported the cooling of nanomechanical [2] and micromechanical [3] oscillators to their lowest energy state using optical forces in a high photon population regime ($10^3$-$10^4$ photons).

High optical quality factors combined with small modal volumes make this kind of devices prone to thermal effects, which strongly depend on the intensity of the electromagnetic field. Indeed, thermo-optic effects are commonly observed in the testing of nano/micro-photonic cavities [4-8] but a combined study of the modification of the optical and mechanical modes in optomechanical (OM) devices is, to the best of our knowledge, still missing. In the present work we address those modifications in a 1D silicon OM crystal cavity as a function of the intracavity photon number when an effective increase of the temperature of the cavity region is the dominant effect.

## II. DESCRIPTION OF THE OM CRYSTALS UNDER STUDY

The fabricated structures are 1D silicon OM crystals whose geometry is similar to that used by Eichenfield *et al.* [9] (Fig. 1a). The defect region consists of 16 central holes in which the hole-to-hole spacing (pitch) is varied quadratically from the nominal lattice constant ($\Lambda$=362 nm) at the beam perimeter to 85% of that value for the holes in the central part. The pitch is kept constant on both sides of the defect over 30 cells, acting as an effective mirror for the the modes confined in the defect region, so that the total number of cells is 76.

The structures were fabricated in Silicon-on-Insulator (SOI) samples with a top silicon layer thickness of 220 nm (resistivity $\rho$ ~1–10 [Ohm cm], p-doping of ~1×$10^{15}$ [$cm^{-3}$]) and a buried oxide layer thickness of 2 [μm]. The OM crystal cavities fabrication process was based on electron beam direct writing on a coated 170 nm of PMMA 950K resist layer. The electron beam exposure was optimized with an acceleration voltage of 10 [KeV] and an aperture size of 30 [μm] with a Raith150 tool. After developing, the resist patterns were transferred into the SOI samples by inductively coupled plasma- reactive ion etching. In order to release the silicon membranes, a second photolithography process based on UV exposure with a mask-aligner tool was employed to open a window in the UV resist just where the structures were placed. Finally, the silicon dioxide under the membranes was removed by using a BHF bath.

This particular design allows the existence of five TE polarized photonic eigenmodes with different energy and spatial distribution along the long axis of the OM structure (see Fig. 1b for the case of the lowest order mode, i.e. the one at the lowest resonant wavelength ($\lambda_{str}$)); these have been simulated using a finite-difference time-domain (FDTD) optical package. [10]

**III. EXPERIMENTAL SETUP**

The experiments were made in a standard set-up for characterizing optical and mechanical properties of OM crystals. A tuneable infrared laser covering the spectral range 1460-1580 [nm] was connected to a tapered fibre with a microloop shape [11]. The polarization state of the light entering the tapered region was set with a polarization controller. The fibre was brought into contact with the etched frame, while the thinnest part of the fibre passes above the central region of the OM crystal. The long tail of the evanescent field (several hundreds of nanometers) and the relatively good spatial resolution (~5 [$\mu m^2$]) of the tapered fibre allowed the local excitation of resonant optical modes of the OM crystal. To enable this configuration and maximize the excitation localization, the fibre formed an angle of about 30º with respect to the OM structure. The gap distance between the fibre and the OM crystal was decreased by sliding the contact point towards the frame edge and increased by doing the opposite (Fig. 1c). A polarization analyser was placed after the tapered fibre region.

The OM crystal photonic cavity is a bi-directional one, which can decay in both forward and backward propagating fibre modes. Only the forward channel was detected by measuring the transmitted optical power ($P_{out}$). The transmission spectra were collected by tuning the laser wavelength ($\lambda_{laser}$) from short to long wavelengths. To check for the presence of a radiofrequency (RF) modulation of the transmitted signal we used an InGaAs fast photoreceiver with a bandwidth of 12 [GHz]. The RF voltage was connected to the 50 [Ohm] input impedance of a signal analyser with a bandwidth of 13.5 [GHz].

All the measurements were performed in an anti-vibration cage at atmospheric conditions of air pressure and temperature.

## IV. OPTOMECHANICAL CHARACTERIZATION

The optical spectrum extracted from the device under study is shown in Fig. 2a. The spectral range covered by the tuneable laser allowed the observation of the first three localized modes supported by the structure, which are polarized in the plane of the sample (TE polarization). The oscillations outside the resonant frequencies reflect the presence of whispering gallery modes in the microlooped tapered fibre. At low input powers, the spectral width ($\delta\lambda$) of the localized modes is about 0.03 nm, which translates into maximum optical quality factors ($Q_o=\lambda_{str}/\delta\lambda$) of 54000 (inset of Fig. 2a). It is worth noting that $Q_o$ can be also written as $Q_o=2\pi c/(\kappa\lambda_{str})$, where $\kappa$ is the overall decay rate of the stored optical energy resulting of the combination of the intrinsic and extrinsic decay rates ($\kappa_i$ and $\kappa_e$ respectively). It is possible to extract $\kappa_i$ and $\kappa_e$ independently by measuring the transmitted fraction, which in the case under study is $P_{out}/P_{in}=(1-\kappa_e/\kappa)^2$ [12].

In the following, we focus on the first optical mode of the OM crystal cavity, which has the lowest modal volume ($V_m \sim 1.4(\lambda_{str}/n)^3$, $n$ being the refractive index) and the highest $Q_o$. These properties allow the highest intracavity photon number ($n_{ph}=2 P_{in} (\kappa_e/\kappa^2)\lambda_{str}/hc$) for a given input laser power ($P_{in}$) and, thus, the highest averaged photon density ($n_{ph}/V_m$). Under such conditions we obtained the RF spectrum plotted in the bottom panels of Fig. 2. The OM crystal presents strong transduced signal for a set of three mechanical modal families, denoted by 'pinch' (Fig. 2b), 'accordion' (Fig. 2c) and 'breathing' (Fig. 2d) modes in order of increasing frequency [9]. Within these families, there is a rich substructure of different mechanical modes associated with different modal orders. A maximum mechanical quality factor ($Q_m$) of 1050 was measured in the highest energy region, probably dominated by scattering with thermal phonons.

## V. TUNING THE OPTICAL AND MECHANICAL MODES OF THE OPTOMECHANICAL CRYSTAL WITH THE PHOTON NUMBER.

Since $\lambda_{laser}$ is around half the band gap of Si, the main mechanisms of light absorption within the OM crystal cavity is two-photon absorption (TPA) [4]. As a consequence, the free-carrier density ($N$) is increased in the optical cavity, reducing the effective refractive index of the optical mode, i.e., causing a blue shift of $\lambda_{str}$, through free-carrier-dispersion (FCD) and introducing further light absorption through free-carrier absorption (FCA) [13]. The free-carriers will finally recombine due to surface recombination (SR). On a stationary situation $N$ can be expressed in the following way [14]:

$$N = \tau_{SR}\beta\left(\frac{hc}{\lambda_{str}}\right)\frac{c^2}{n^2}\left(\frac{n_{ph}}{V_m}\right)^2 \quad (1)$$

, where $\tau_{SR}$ is the free-carriers effective lifetime, which also accounts for the diffusion, and $\beta$ is the TPA coefficient. Typical values for bulk silicon are $\tau_{SR}$~1 [ns] and $\beta$~0.8 [cm/GW] [4]. The highest $n_{ph}$ achieved in this work was on the order of $10^4$, which means that $N$ was always below $10^{17}$ [cm$^{-3}$]. It is important to note that, although $N$ follows the spatial profile of the optical mode, in Eq. (1) it is assumed to be homogeneous in the cavity region. Among the different free-carrier relaxation mechanisms leading to an effective temperature increase of the cavity region ($\Delta T$) we only consider the FCA induced intraband relaxation and the interband recombination, which is mediated by the surface states. Thus, in the stationary situation, $\Delta T$ can be written as:

$$\Delta T = \tau_{TH} N \left( \alpha_{FCA}\left(\frac{n_{ph}}{V_m}\right) + \alpha_{SR} \right) \quad (2)$$

, where $\alpha_{FCA}$ and $\alpha_{SR}$ are defined as the FCA and SR thermal rates respectively and $\tau_{TH}$ is the $\Delta T$ characteristic lifetime. It has to be noted that the exact determination of the spatial temperature distribution within the OM crystal for a given $n_{ph}$ is an extremely complex problems. In fact, to start with, it depends on the effective thermal conductivity of the resonator, which is significantly reduced with respect to the bulk values below a slab-layer thickness of few micrometers [16] and is affected by the specific geometric nanostructure of the Si beam as well [17], [18]. In addition, the heat source is not a point one, since it extends over the cavity volume following the specific optical mode spatial profile.

The refractive index increase due to $\Delta T$ is known as the thermo-optic (TO) effect, which shifts $\lambda_{str}$ towards longer wavelengths. The overall resonant wavelength drift ($\Delta\lambda$) associated to TO and FCD effects can be expressed in terms of a power series of $N$ and $\Delta T$. By keeping the linear terms this is written as:

$$\Delta\lambda = -\frac{\partial \lambda_{str}}{\partial N} N + \frac{\partial \lambda_{str}}{\partial T}\Delta T. \quad (3)$$

The FCD and the TO coefficients for this particular structure were calculated to be $\frac{\partial \lambda_{str}}{\partial N}$=6.8x10$^{-19}$ [nm cm$^3$] and $\frac{\partial \lambda_{str}}{\partial T}$=5.6x10$^{-2}$ [nm K$^{-1}$] respectively (more details of these calculations will be given below). It is worth noting we have neglected Kerr effects with respect to the FCD and TO contributions [4].

The set of spectra of Fig. 3a, which were obtained for different $P_{in}$, shows that $\lambda_{str}$ experiences a red-shift that increases with $n_{ph}$. Redshifts up to 70 linewidths ($\Delta\lambda$=2.1 [nm]) are observed since the overall TO contribution largely prevails over FCD. Similar trends with $P_{in}$ have been reported in literature for other types of optical resonators [4-7]. While $\lambda_{laser}<\lambda_{str}$ there is a red-shift of $\lambda_{str}$ that increases during the $\lambda_{laser}$ sweep. The transmission minimum occurs when $\lambda_{laser}=\lambda_{str}$, i.e. the light absorption is maximum and, consequently, so it is the red-shift. For higher values of $\lambda_{laser}$ the absorption decreases and the cavity relaxes back to the initial cold state, leading to the abrupt transmission jump.

The resonance contrast becomes smaller with increasing $P_{in}$ as a consequence of a decrease of $Q_o$. This is related to an increase of $\kappa_i$ due to FCA losses (Fig. 3b), which are directly proportional to $N$ [4]. Although $\kappa_i$ has been approximated to be spatially homogeneous, it actually depends on the free carrier spatial distribution, whose behavior with $n_{ph}$ is quadratic. The data of Fig. 3b suggest that the intrinsic decay rate at low $n_{ph}$ ($n_{ph}<2\times10^2$), i.e., when the FCA contribution is negligible, is about $\kappa_i/2\pi$=2.9 [GHz].

The determination of the TO and the FCD coefficients is done by assuming that the observed $\Delta\lambda$ is only associated to an average change in the Si refractive index ($n_{Si}$) of the set of cells defining the defect region, i.e., between the two mirrors. A linear dependence of the resonance optical frequency on $n_{Si}$ is extracted from FDTD optical simulations. The TO coefficient is then calculated by combining the latter result with the well-known linear relation of $n_{Si}$ with temperature ($T$), $n_{Si} = 3.38\cdot(1+3.9\times10^{-5}[K^{-1}]T[K])$ at $\lambda_{str}$=1.55 [μm] [19]. In an equivalent way, the FCD coefficient is extracted using the relation of $n_{Si}$ with $N$, $n_{Si} = 3.42 + 1.35\times10^{-21}[cm^3] N[cm^{-3}]$ at $\lambda_{str}$=1.55 [μm] [20]. It is possible to calculate the FCD contribution to $\Delta\lambda$, i.e., $-\frac{\partial\lambda_{str}}{\partial N}N$, for a given $n_{ph}$ by determining $N$ using Eq. (1). Using Eq. (3), the TO contribution to $\Delta\lambda$ is finally estimated, being one order of magnitude greater than the FCD counterpart. Thus, it is straightforward to establish a relation between $\Delta\lambda$ and $\Delta T$. A superlinear response of $\Delta T$ with respect to $n_{ph}$ is found (Fig. 4), which is consistent with Eq. (2).

We have also observed that the spectral shift achieved when exciting higher order optical modes (slightly lower $Q_o$ values and higher $V_m$) is slightly reduced with respect to the first mode, even though $n_{ph}$ can be forced to be similar by increasing $P_{in}$. In fact, the temperature spatial distribution is different for each optical mode and $\Delta T$ decreases with the reduction of $n_{ph}/V_m$ (see Eqs. 1 and 2).

In addition to the optical resonance tuning, there is also a shift in the frequencies of the mechanical modes ($\Omega_o$) when changing $n_{ph}$ in the OM crystal. This was achieved by changing $\lambda_{laser}$ at high transmitted powers whilst keeping the resonant condition. Given the low $\Omega_o/\kappa$ values of our structures (always in the unresolved regime, i.e., $\Omega_o/\kappa<1$, even for the lowest values of $n_{ph}$), dynamical back-action effects are not detected and $Q_m$ remains almost invariant. In the case illustrated in Fig. 5 the maximum optical resonance drift was 10 [nm] (≈300 linewidths) associated to $n_{ph} = 2\times10^4$. In this extreme condition we have observed a 0.3% frequency downshift (about 7 [MHz]) in the 'breathing' modes. Slightly lower relative shifts were observed for the modes present in lower frequency regions (insets of Fig. 5). The reported frequency reduction of the mechanical modes is a consequence of the decrease of the Si Young's modulus, which is related to the effective temperature increase reported above [21, 22]. Again, since there is a lack of knowledge of the temperature spatial distribution within the OM crystal, it is not possible to extract the spatial distribution of the Young's modulus. Nevertheless, finite-element method (FEM) simulations using the homogeneous increase of the cavity temperature extracted from the optical resonance shift, i.e., a homogeneous reduction of the Young's modulus in the cavity region, have provided frequency downshifts compatible with those experimentally reported (green dashed line of Fig. 5). The observed optical and mechanical shifts are compatible and well described with a TO contribution. The small deviation of the experimental mechanical spectra from the expected linear behaviour is associated to optical-spring effects, which tend to increase the frequency of the mechanical eigenmodes.

FEM simulations reveal that, while the eigenfrequency shift of the 'breathing' and 'accordion' modes accounts for the temperature change of the whole defect region, the drift of a 'pinch' mode is dominated by the temperature increase of the volume of the 2-3 cells where the mode is highly localized. Interestingly, the frequency of each 'pinch' mode increases as it is localized further from the center of the structure (manifold of Fig. 2b). On the other hand, when the maximum pitch decreasing is high enough, the amount of modes within that family roughly scales with the number of cells composing the defect region. Those particular features could be engineered for extracting high resolution temperature spatial profiles in OM crystal cavities, where the spatial range will be given by the number of cells defining the defect and the resolution by the effective spatial separation between consecutive modes. In the specific case presented here, the low number of defect cells forces the fundamental optical mode to fill the defect region, so the flat temperature assumption already provides a fair description of the 'pinch' modes shift. In fact, the spectral shift difference among the observed 'pinch' modes is not meaningful.

**VI. CONCLUSIONS**

We have presented a study of the optical and mechanical properties of a silicon 1D OM crystal cavity in relation with the temperature increase associated to the absorption of part of the optical energy stored in the cavity. The relatively high optical quality factor of the supported modes allows storing photon numbers in the range of $10^4$ in modal volumes of the order of $(\lambda_{str}/n)^3$ using input powers below 1 [mW]. This leads to the activation of thermo-optic effects that provoke a shift towards longer wavelengths on the optical modes and towards lower frequencies on the mechanical modes. An effective reduction of the silicon Young's modulus associated to the temperatures extracted from the optical measurements provide a reasonably good description of the behaviour of the mechanical modes. The exact calculations would involve the determination of the spatial profile of the temperature increase associated to the absorption of light of a particular eigenmode, which is an extremely challenging problem. We propose the use of 'pinch' modes in OM crystals with a high number of cells in the defect region for extracting temperature spatial profiles with high resolution. We believe that these results shed more light on the understanding of silicon based OM crystals in a high photon number regime and will help in their further control and development.

**ACKNOWLEDGEMENTS**

This work was supported by the EU through the project TAILPHOX (ICT-FP7-233883) and the ERC Advanced Grant SOULMAN (ERC-FP7-321122) and the Spanish projects TAPHOR (MAT2012-31392). The authors thank A. Tredicucci for a critical reading of the manuscript and A. Pitanti for fruitful discussions.

**FIGURES**

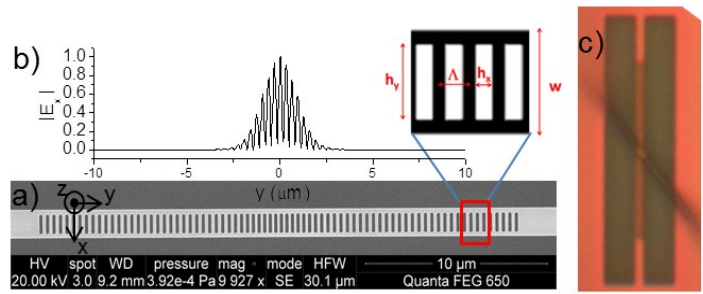

Fig. 1. a) SEM micrograph of one of the fabricated OM crystals. The geometrical parameters are $\Lambda$=362 [nm], $w$=1396 [nm], $h_y$=992 [nm], $h_x$=190 [nm] and thickness=220 [nm]. b) |$E_x$| along the y axis for the fundamental optical mode. c) Top view of the relative positioning of the tapered fibre and the OM crystal as seen with a 50X optical microscope. The fibre passes close enough to the central part of the OM crystal to excite its localized photonic modes.

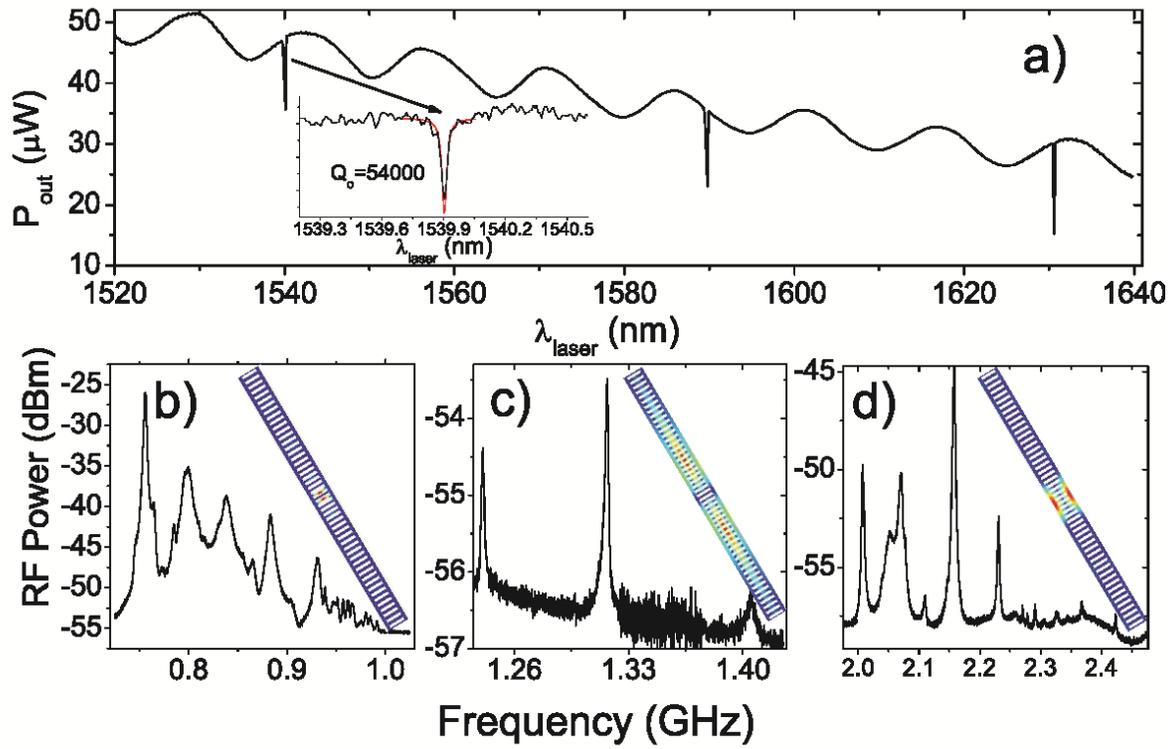

Fig. 2. Optical (panel a)) and RF (bottom panels, log scale in vertical) spectra of the OM crystal. The RF spectrum is separated in three panels corresponding to the different modal families supported by the structure. Panel b), c) and d) correspond to the 'pinch', 'accordion' and 'breathing' families respectively. The simulated deformation profiles of the first mode of each family are also included.

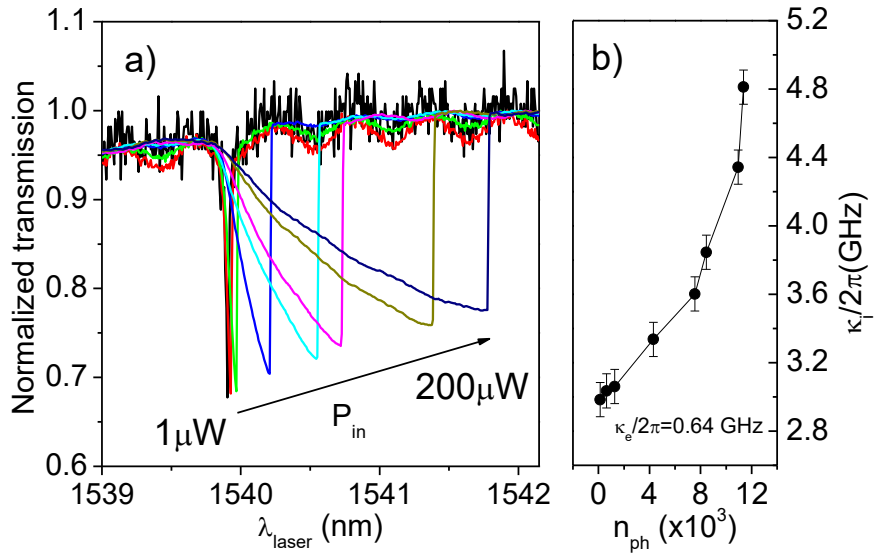

Fig. 3. a) Normalized spectra of the first optical mode obtained for different $P_{in}$. b) Intrinsic decay rate as a function of the photon number. The extrinsic $n_{ph}$ decay rate is $\kappa_e/2\pi$=0.64 [GHz].

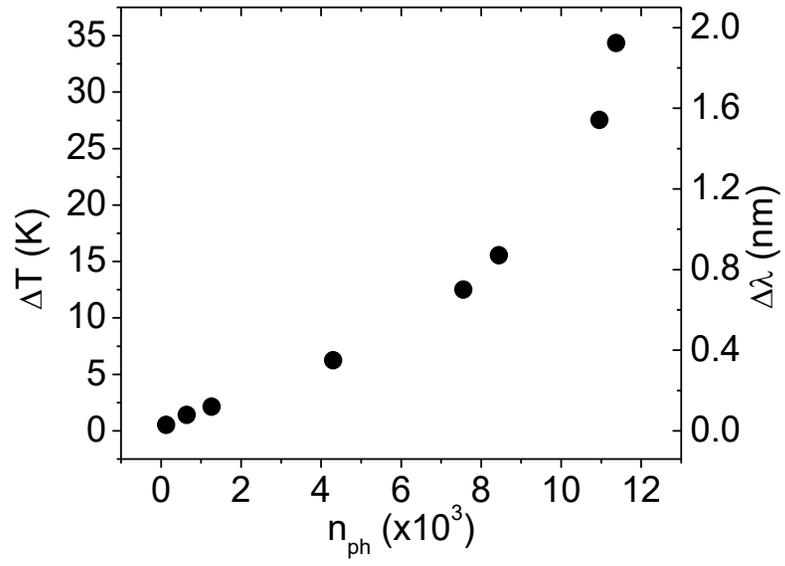

Fig. 4. Effective temperature increase (wavelength drift in the right axis) as a function of the intracavity photon number.

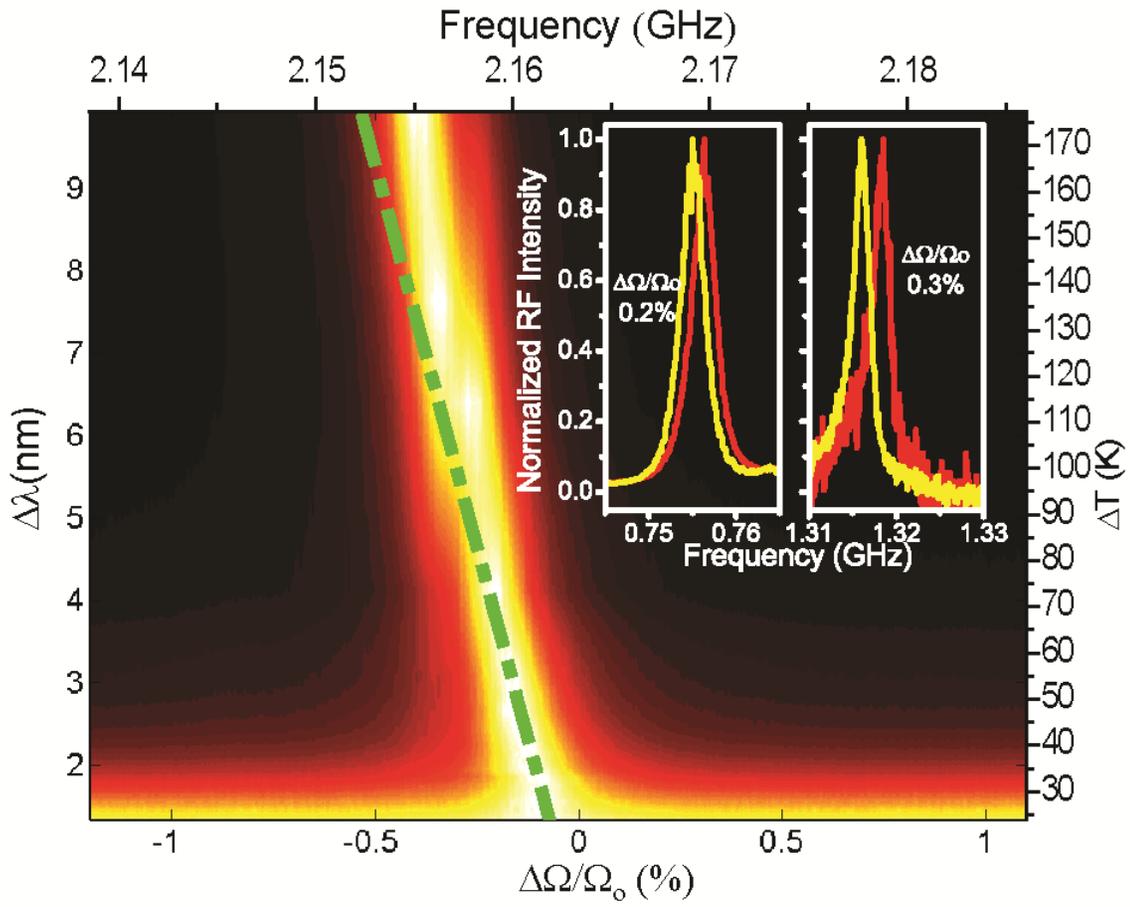

Fig. 5. a) Colour contour plot of the normalized RF power (in linear scale) of the first 'breathing' mode as a function of the relative frequency detuning (x axis) and the optical wavelength (y axis). The photon number reaches $n_{ph}=2\times10^4$ for the strongest drift. The dashed line represents the FEM simulation using the homogeneous increase of the cavity temperature extracted from the optical resonance shift. Insets. 'Pinch' (left panel) and 'accordion' (right panel) modes at the extreme optical wavelengths.